\documentclass[a4paper,12pt]{article}
\pdfoutput=1

\usepackage[utf8]{inputenc}
\usepackage[T1]{fontenc}
\usepackage{dblfloatfix} % to use modifiers [h] etc. in \begin{figure*} ... \end{figure*}
\usepackage{fixltx2e} % prevent starred-figures from appearing out of ascending order
\usepackage{float}
\usepackage{wrapfig}
\usepackage{rotating}
\usepackage[normalem]{ulem}
\usepackage[font=small,labelfont=bf,labelsep=endash,margin=20pt]{caption}
\usepackage[title,header]{appendix}

\usepackage{relsize} % for relative font sizes, useful in commands such as PCa, Wnt
\usepackage{amsmath}
\usepackage{marvosym}
\usepackage{wasysym}
\usepackage{amssymb}
\usepackage{latexsym}
\usepackage{physics}

\usepackage{bm}
\usepackage{mathtools}
\usepackage{newtxtext,newtxmath}
\usepackage{isomath} % italicize upper case Greek
\usepackage[margin=25mm]{geometry}
\usepackage{grffile}% allow spaces in graphicx filenames
\usepackage{enumitem} %\begin{enumerate}[itemsep=2pt,parsep=2pt,topsep=0pt,partopsep=0pt]
\setlist{itemsep=2pt,parsep=2pt,topsep=0pt,partopsep=0pt}
\usepackage{graphicx}
\usepackage{url}
\usepackage[hyperref,x11names]{xcolor}
\usepackage[colorlinks=true,linkcolor=Firebrick4,citecolor=blue,urlcolor=SteelBlue4]{hyperref}
\usepackage{tikz}
\usepackage{pgfplots}
\pgfplotsset{compat=1.15}
\usepackage{ctable} % \toprule, \thickmidrule, \midrule, \cmidrule and \bottomrule to replace \hline in tables
\usepackage{tabularx} % For extensible columns. Alternative: tabulary.
\usepackage{cite} % to display references [1, 2, 3] as [1-3] (without using natbib)
\usepackage{mathbbol} % \mathbb{0} .. \mathbb{9}
\usepackage{bbm} % \mathbbm{0} .. \mathbbm{9}, \mathbbm{R} etc.
\usepackage{thmtools} % to have boxed theorems and similar type environments
\usepackage{siunitx}

\renewcommand{\d}{\mathrm{d}}

\newcommand{\p}{\partial}
\newcommand{\pd}[2]{\frac{\partial #1}{\partial #2}}

\newcommand{\Order}{\mathrm{O}}
\newcommand{\e}{\mathrm{e}}

\renewcommand{\b}[1]{{\bm{#1}}} % usually italic bold (greek letter etc.) [new]

\renewcommand{\geq}{\geqslant}

\newcommand{\keywords}[1]{\vspace{2mm}\noindent\textbf{Key words:} #1} % use within abstract

\let\paragraphold\paragraph
\renewcommand*{\paragraph}[1]{\paragraphold{#1.}} % automatically add a dot after argument of \paragraph

\def\cbeg{}
\def\cend{}

\begin{document}
\title{\bf Curvature dependences of wave propagation in reaction--diffusion models}
\renewcommand{\thefootnote}{\fnsymbol{footnote}}%
\author{Pascal R Buenzli$^\text{a,}$\footnotemark[1]\ , Matthew J Simpson$^\text{a}$}

\date{\small \vspace{-2mm}$^\text{a}$School of Mathematical Sciences, Queensland University of Technology (QUT), Brisbane, Australia\\\vskip 1mm \normalsize \today\vspace*{-5mm}}

\maketitle
\begin{abstract}
     Reaction--diffusion waves in multiple spatial dimensions advance at a rate that strongly depends on the curvature of the wave fronts. These waves have important applications in many physical, ecological, and biological systems. In this work, we analyse curvature dependences of travelling fronts in a single reaction--diffusion equation with general reaction term. We derive an exact, non-perturbative curvature dependence of the speed of travelling fronts that arises from transverse diffusion occurring parallel to the wave front. Inward-propagating waves are characterised by three phases: an establishment phase dominated by initial and boundary conditions, a travelling-wave-like phase in which normal velocity matches standard results from singular perturbation theory, and a dip-filling phase where the collision and interaction of fronts create additional curvature dependences to their progression rate. We analyse these behaviours and additional curvature dependences using a combination of asymptotic analyses and numerical simulations.

     \keywords{Excitable media, wave front, travelling wave, moving boundary problems, invasion}
\end{abstract}

\protect\footnotetext[1]{Corresponding author. Email address: \texttt{pascal.buenzli@qut.edu.au}}%
\renewcommand{\thefootnote}{\arabic{footnote}}%

\section{Introduction}
Travelling fronts in reaction--diffusion models describe the progression in space of the transition between two states of an excitable system. These fronts represent the advance of a wave of excitation, where spatial locations in a rest state transition into an excited state~\cite{tyson-keener-1988,grindrod-lewis-murray-1991}. Examples of such propagating waves abound in physical, biological, and ecological systems. Particular examples of these waves are electrical activity along axonal membranes, chemical activity in chemical reactions, flame propagation in wildfires, biochemical activity in multicellular systems, biological tissue growth, biological species invasion, infection disease spread, social outbursts, as well as crystal growth and star formation in galaxies~\cite{tyson-keener-1988,foerster-mueller-hess-1988,grindrod-lewis-murray-1991,lewis-kareiva-1993,Echekki1996,volpert-petrovskii-2009,hilton-etal-2016,Dave2020,buenzli-etal-2020,bakhshi-etal-2021}.

Wave propagation in these systems results from the combination of diffusion and autocatalytic production of the excitable element. In one spatial dimension, excitable media support the establishment of travelling waves with constant propagation speed $c$ in the long-time limit, under appropriate initial and boundary conditions~\cite{murray-2}. In higher spatial dimensions, it is well known experimentally and theoretically, that the propagation speed of travelling fronts is modified by the front's curvature, which in general is a function of space and time~\cite{tyson-keener-1988,foerster-mueller-hess-1988,witelski-etal-2000,volpert-petrovskii-2009,buenzli-etal-2020}. Singular perturbation theories of excitable reaction--diffusion systems show that the normal velocity $v$ of travelling fronts is given by
\begin{align}\label{v-asymptotic}
    v = c - D\kappa
\end{align}
at leading order in the diffusivity $D>0$ as $D\to 0$, where $\kappa$ is the front's mean  curvature~\cite{tyson-keener-1988,grindrod-lewis-murray-1991,lewis-kareiva-1993}.

In principle, Eq.~\eqref{v-asymptotic} enables the determination of the spatial location of the travelling front as a function of time from an initial condition. It is referred to as the ``eikonal equation for reaction--diffusion systems'' for this reason, by analogy to geometrical optics. Travelling front velocities are used in numerical marker particle methods as well as level set methods applied to moving boundary problems~\cite{sethian-1999,osher-fedkiw-2003,garcia-reimbert-etal-2002,leung-zhao-2009,hon-leung-zhao-2014}. Several results are known about the behaviour of fronts moving with curvature-dependent speeds. Curve-shortening flows and mean curvature flows for which normal velocity is proportional to curvature, as in Eq.~\eqref{v-asymptotic}, round off the initial front shape in such a way that the front becomes circular before shrinking into a point~\cite{brakke-1978,grayson-1987,sethian-1999,chou-zhu-1999,liu-2013}. Such rounding of initial shape is commonly observed in experimental systems~\cite{foerster-mueller-hess-1988,rumpler-etal-2008,hilton-etal-2016,buenzli-etal-2020}. Many variants of mean curvature flows, such as lengh-preserving flows, area-preserving flows, and Willmore flows are proposed as effective models of the evolution of fronts or interfaces~\cite{chen-giga-goto-1991,sethian-1999,osher-fedkiw-2003,giga-2006,liu-2013,dallaston-mccue-2016,andrews-etal-2020}. There is recent interest in elucidating curvature-dependent velocities in moving fronts that represent the growth of biological tissues due to their application in tissue engineering~\cite{rumpler-etal-2008,bidan-etal-2012,guyot-etal-2015,wang-etal-2018,mccue-etal-2019,buenzli-etal-2020,callens-etal-2020,browning-etal-2021}. In this context, the eikonal equation represents an understanding of a biological growth law based on migratory and proliferative cellular behaviours~\cite{goriely-2017,alias-buenzli-2017,alias-buenzli-2018,alias-buenzli-2019,hegarty-cremer-etal-2021}.

For the eikonal equation~\eqref{v-asymptotic} to be a useful representation of the evolution of excitable systems, it is important that it represents the normal speed of travelling fronts accurately. Equation~\eqref{v-asymptotic} is derived rigorously under mathematical assumptions on solution profiles. However, it is unknown how well these assumptions hold in practice. The plane-wave travelling speed $c$ is proportional to $D^{1/2}$~\cite{tyson-keener-1988,murray-2}, so the curvature-dependent term in Eq.~\eqref{v-asymptotic} is only relevant for curvatures of order $\Order(D^{-1/2})$. This corresponds to slowly moving highly curved fronts, such as where circular wave fronts collide and lead to an accelerating phase in species invasion fronts~\cite{tyson-keener-1988,volpert-petrovskii-2009}. Such colliding circular fronts, however, can be expected to not conform to the assumptions used to derive Eq.~\eqref{v-asymptotic} at some point. It is also unclear how boundary conditions may impact the eikonal equation for fronts located near boundaries of the domain.

In this work, we investigate the eikonal equation describing the normal speed of travelling fronts non perturbatively in $D$. For simplicity, we consider a reaction--diffusion system consisting of a single species only, with linear diffusion, and an arbitrary reaction term. We perform numerical simulations of inward and outward wave propagation in domains of different shapes. This allows us to initiate fronts of the solution with various curvature, including shapes with positive and negative curvatures. We also investigate the influence of different boundary conditions imposed on the solution at the domain boundaries.

Our numerical simulations suggest that Eq.~\eqref{v-asymptotic} is only valid in practice in a restricted time window, away from boundaries, and before circular fronts collide. We show numerically and analytically that there are several contributions to the curvature dependence of the normal speed of travelling fronts. The linear dependence upon curvature in Eq.~\eqref{v-asymptotic} is an exact, non perturbative contribution that originates from the transverse component of diffusion (diffusion parallel to the wave front). Further dependences upon curvature, including one of the same order in diffusivity, are exhibited explicitly using large-curvature asymptotic analyses and linear models. These further dependences arise from the normal component of diffusion and from autocatalytic production of the species in a spatially restricted environment.

\begin{figure}
        \centering\includegraphics[width=0.95\textwidth]{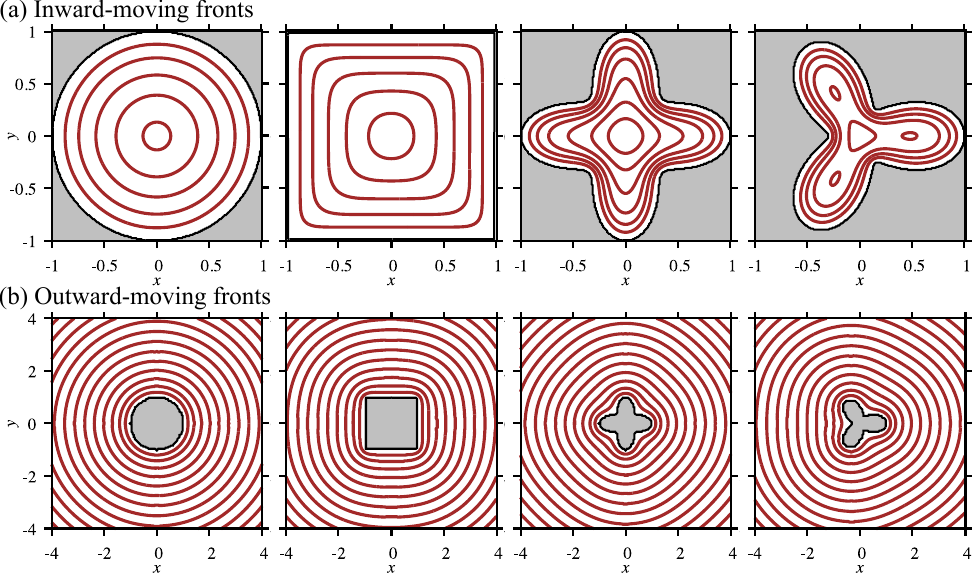}
        % inward: D=0.005, lambda=1, L=2
        % outward: D=0.02, lambda=1, L=15
        \caption{Numerical simulations of the Fisher--KPP reaction--diffusion equation ($F(u) = \lambda u(1-u)$) on domains of different shapes (white) and linear size $L$, showing that the speed of travelling fronts $u(\b r, t)=u_c$ (red) is affected by the front's curvature; fronts $u_c=0.5$ are shown every 1.5 time units, except for the inward-moving fronts of the cross and of the petal which are shown every 0.7 time units. (a) Inward-moving travelling fronts with $\lambda=1$, $D=0.005$; (b) Outward-moving travelling fronts with $\lambda=1, D=0.02$. In all these simulations, $u(\b r, 0)=0$ in the domain and $u(\b r, t)=1\ \forall t$ on the domain boundary (black)}
        \label{fig:fronts}
\end{figure}

\begin{figure}
        \centering\includegraphics{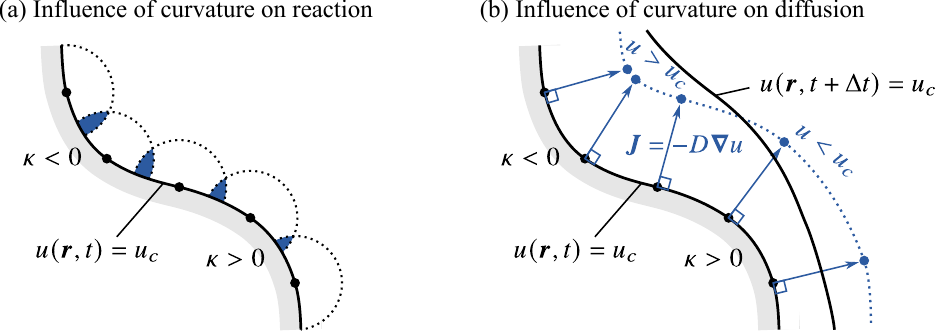}
        \caption{The curvature of contours $u=u_c$ (solid black lines) influences contour propagation speed via both (a) reaction rates and (b) diffusion. (a) The probability of particles (black dots) undergoing a reaction is enhanced at concavities ($\kappa <0$) because particles are more likely to encounter there as they accumulate by autocatalytic production, and it is reduced at convexities ($\kappa >0$) because particles are less likely to encounter there, as indicated by the blue shaded areas. (b) Particles (black dots) evenly distributed on a constant density contour $u(\b r, t)=u_c$ at time $t$ move on average with diffusive flux $\b J=-D\grad u$ normal to the contour in the direction of lower density (blue arrows). Near concavities, particles move closer to each other, which increases their density ($u>u_c$). Near convexitites, particles move away from each other, which decreases their density ($u<u_c$). The constant density contour $u(\b r, t+\Deltaup t)=u_c$ at time $t+\Deltaup t$ is thereby pushed further ahead near concavities than near convexities.}
        \label{fig:conceptual}
\end{figure}

\section{Model description and theoretical developments}

We consider the single-species reaction--diffusion model
\begin{align}\label{reac-diff}
    u_t = D\laplacian u + F(u),
\end{align}
describing the spatio-temporal evolution of a quantity $u(\b r, t)$, where $\b r$ is a position vector and $t$ is time, with constant diffusivity $D>0$ and general reaction term $F(u)$. The subscript $u_t$ denotes $\p u/\p t$, and $\laplacian u$ is the Laplacian of $u$. The quantity $u$ may represent the density of entities such as a chemical species, particles, or cells forming a biological tissue, that undergo diffusive motion, and that may be generated and eliminated through the source term $F(u)$. While the exact curvature dependence elucidated below is valid in two- and three-dimensional space, our specific examples and further analyses will focus on two-dimensional domains~$\Omega$ for simplicity. We supplement Eq.~\eqref{reac-diff} with either inhomogeneous Dirichlet boundary conditions, or homogeneous Neumann (no flux) boundary conditions:
\begin{align*}
  &u\vert_{\p\Omega} = u^*\quad \text{(Dirichlet)}
    \qquad\text{or}\qquad \b n \vdot \grad u\vert_{\p\Omega} = 0\quad  \text{(Neumann)},
\end{align*}
where $\b n$ is the inward-pointing unit normal vector to the boundary $\p\Omega$ of the domain. We assume the domain to be initially empty and set $u(\b r, 0) = 0$ in the interior of $\Omega$. With Dirichlet boundary conditions, we set $u(\b r, 0) = u^*$ at the boundary $\p\Omega$. With Neumann boundary conditions, we set $u(\b r, 0) = u^*\deltaup_{\p\Omega}(\b r)$, where $\deltaup_{\p\Omega}$ is a surface delta distribution~\cite{buenzli-2016}. Formally, $u$ is zero everywhere except at the boundary $\p\Omega$ where $u$ is infinite, such that $\int\d\b r\  u(\b r, 0)=\int_{\p\Omega}\d\sigma = u^*|\p\Omega|$ is proportional to the perimeter of $\p\Omega$.

A source term $F(u)$ of particular interest is the logistic growth term $F(u)=\lambda u(1-u)$ for a normalised density $u\in [0,1]$, in which case Eq.~\eqref{reac-diff} is the two-dimensional analogue of the Fisher--KPP model~\cite{murray-2,edelstein-keshet}. In this model, $u=0$ represents the unstable, excitable rest state, and $u=1$ represents the stable, excited state. In infinite space and under appropriate initial and boundary conditions, this model is well-known to lead to travelling-wave solutions where rest states progressively transition into excited states~\cite{murray-1,murray-2,edelstein-keshet}.  However, we do not assume necessarily that the source term $F(u)$ leads to travelling-wave solutions, nor that it possesses rest states and excited states. We will state results valid for general source terms $F(u)$, and will illustrate properties of fronts of the solution to Eq.~\eqref{reac-diff} with linear source terms, $F(u)=\lambda u$ and $F(u)=\lambda(1-u)$. These linear models do not support travelling-wave solutions. The time-dependent speed of their travelling fronts enable us to illustrate important properties of the eikonal equation. We also consider a bistable source term $F(u)=\lambda u(u-a)(1-u)$ with $0<a<1$, a common model for population growth subject to a strong Allee effect. The rest state $u=0$ and excited state $u=1$ in this model are both stable, yet this model still exhibits travelling-wave transitions from $u=0$ to $u=1$ or from $u=1$ to $u=0$ depending on $a$ and on the shape and size of initial conditions~\cite{lewis-kareiva-1993,Li2022}.

\paragraph{Speed of travelling fronts}
A travelling front of $u(\b r, t)$ is defined to be the time-dependent set of points $\b r$ in the domain $\Omega$ such that $u(\b r, t)=u_c$ for a constant value $u_c$ (Figure~\ref{fig:fronts}). A convenient way to estimate the velocity of travelling fronts is to calculate the velocity of contours of $u$ at each location of the domain. Each location $\b r$ of the domain has a density $u(\b r, t)=u_c$ for a certain allowable value $u_c$ of the solution. The velocity of a contour $u_c$ at $\b r$ can be determined as in level-set methods~\cite{sethian-1999,osher-fedkiw-2003}. In short, given a parametrisation $s\mapsto \b r(s,t)$ of a contour line at time $t$, the velocity of a point $\b r(s,t)$ of the contour at fixed $s$ is $\b r_t$. Differentiating $u\left(\b r(s,t), t\right)) = u_c$ with respect to $t$, and using the fact that the unit normal to level sets of $u$ in the decreasing direction of $u$ is $\b n=-\grad u/|\grad u|$, one obtains
\begin{align*}
    u_t + \grad{u}\vdot \b r_t = u_t - |\grad u|\b n\vdot r_t = u_t - |\grad u| v = 0,
\end{align*}
where $v=\b n\vdot \b r_t$ is the normal speed in the direction of lower values of $u$, by definition of $\b n$. Given a solution $u(\b r, t)$, the normal speed of travelling fronts can thus be estimated at any point of the domain and any time by evaluating~\cite{Echekki1996}:
\begin{align}\label{v-levelset}
    v = \frac{u_t}{|\grad u|} = \frac{1}{|\grad u|}\left( D\laplacian u + F(u)\right).
\end{align}
It is clear from Eq.~\eqref{v-levelset} that this velocity is, in general, space and time dependent, and that there may be contributions due to both the diffusion term and reaction term of Eq.~\eqref{reac-diff}, as illustrated in Figure~\ref{fig:conceptual}. Equation~\eqref{v-levelset} also describes the speed of travelling-wave solutions that may establish in an infinite domain under appropriate boundary conditions in the limit $t\to\infty$. A travelling-wave solution is such that $u(\b r, t) = u_0(\b r - \b c t)$, where $\b c$ is a constant velocity vector and $u_0$ is the wave profile. For such travelling-wave solutions, $u_t = - \grad u_0\vdot\b c$ and $\grad u = \grad u_0$, so that
\begin{align*}
    v = \frac{u_t}{|\grad u|} = -\frac{\grad u_0\vdot\b c}{|\grad u_0|} = \b n\vdot\b c = c,
\end{align*}
where $\b n$ is the unit vector normal to contours of $u_0$, and $c$ is the normal velocity of the travelling wave. For source terms $F(u)$ and initial and boundary conditions that lead Eq.~\eqref{reac-diff} to establish travelling-wave solutions in the long-time limit, Eq.~\eqref{v-levelset} is such that $\lim_{t\to\infty} v = \b n\vdot\b c = c$. The value of working with Equation~\eqref{v-levelset} is that it provides more practical insight that could be relevant in a finite time, finite space experiment where, strictly speaking, travelling wave solutions never arise.

\paragraph{Exact curvature dependence due to transverse diffusion}
Equation~\eqref{v-levelset} enables us to extract an exact, non-perturbative contribution of the curvature of travelling fronts $u(\b r, t)=u_c$ to their progression rate $v$. In Appendix~\ref{appx:laplacian}, we show that the Laplacian can be decomposed locally into a normal component and a transverse component with respect to the local geometry of the contour $u(\b r, t)=u_c$. The normal component of the Laplacian is $u_{nn} =  \b n^\text{T} H(u)\, \b n = \p^2u/\p n^2$, where $H(u)=\grad{\grad u}$ is the Hessian matrix of $u$ and $\p/\p n = \b n\vdot\grad$. The transverse component of the Laplacian, $\laplacian u - u_{nn}$, is always proportional to the mean curvature $\kappa=\div \b n$ of the travelling front (Appendix~\ref{appx:laplacian}), i.e., $\laplacian u - u_{nn} = - \kappa|\grad u|$, so that
\begin{align}\label{kappa-transverse}
    \laplacian u = u_{nn} - \kappa|\grad u|.
\end{align}
Substituting Eq.~\eqref{kappa-transverse} into Eq.~\eqref{v-levelset} elucidates an exact dependence between the normal velocity of travelling fronts, and their mean curvature:
\begin{align}\label{v}
    v(\kappa) = w - D\kappa,
\end{align}
where
\begin{align}\label{vn}
    w \equiv \frac{1}{|\grad u|}\left[ D u_{nn} + F(u)\right]
\end{align}
is a remaining contribution to $v$ that only depends on the local profile of the solution $u(\b r, t)$ in the normal direction. Indeed, $-|\grad u|=\p u/\p n$ and $u_{nn}$ are derivatives of $u$ in the direction $\b n$.

We note here that our sign conventions for the unit normal vector $\b n$ and curvature $\kappa = \div\b n$ are such that $\kappa < 0$ at concave portions of the domain boundary, and $\kappa > 0$ at convex portions of the domain boundary. This sign convention and interpretation of convexity carries over to travelling fronts $u(\b r, t)=u_c$ inside the domain by interpreting regions where $u>u_c$ as occupied, and regions where $u<u_c$ as empty. With this convention, inward-moving circular fronts of the models $F(u)$ we consider have negative curvature, and outward-moving circular fronts have positive curvature.

The expression $v(\kappa)$ in Eq.~\eqref{v} matches the asymptotic expression~\eqref{v-asymptotic} from singular perturbation theory if $w$ is the speed $c$ of the travelling wave in the corresponding one-dimensional problem on an infinite domain, when such waves exist~\cite{tyson-keener-1988,lewis-kareiva-1993,volpert-petrovskii-2009}. The relationship between $w$ and $c$ can be assessed by comparing Eq.~\eqref{vn} with the speed of travelling fronts in 1D, given from Eq.~\eqref{v-levelset} by
\begin{align}\label{v1D}
    w_\text{1D} = \frac{1}{|u_x|}\left[D u_{xx} + F(u)\right] \stackrel{t\to\infty}{\longrightarrow} c,
\end{align}
where convergence to $c$ holds if the solution becomes a travelling wave in the long-time limit, which typically depends on initial and boundary conditions. The term $-D\kappa$ in Eq.~\eqref{v-asymptotic} is thus not only valid to $\Order(D^2)$ as $D\to 0$, but is an exact dependence, which holds for any value of $D$. While we may have $w \sim w_\text{1D} \sim c$ in some asymptotic limits~\cite{tyson-keener-1988,volpert-petrovskii-2009}, we will show below that $w$ in Eq.~\eqref{v} may also differ significantly from the one-dimensional travelling front speed $w_\text{1D}$ and travelling-wave speed $c$, due to the proximity of boundaries and other geometric constraints affecting the behaviour of the solution in confined spaces. The contribution $w$ in Eq.~\eqref{vn} only involves the profile of $u(\b r, t)$ in the normal direction, but this profile is still coupled via the evolution equation with the solution profile in transverse directions. In other words, $w$ may still have implicit dependences upon the curvature $\kappa$ of the travelling front (which is a transverse property of the front), and these dependences may arise from both reaction and diffusion terms. In the remainder of the paper, we investigate these specific contributions in more detail, through numerical simulations, asymptotic behaviours, and analytic solutions for a range of domain shapes, boundary conditions, and source terms $F(u)$.

\subsection{Numerical simulations}
Numerical simulations of Eq.~\eqref{reac-diff} presented in this work are based on a simple explicit finite difference scheme using forward Euler time stepping, and centred differences in space (FTCS). Accuracy of numerical results is checked by refining computational grids until convergence is observed. Higher order time stepping scheme are possible (such as total variation diminishing Runge-Kutta methods) but curvature effects are known to be more sensitive to spatial discretisation than temporal discretisation~\cite{osher-fedkiw-2003,alias-buenzli-2019}.
% Our choice of an explicit forward Euler scheme is for convenience.

For general domain shapes $\Omega$, we discretise a square domain of size $L\times L$ large enough to contain $\Omega$ using a regular grid with constant space steps $\Deltaup x = \Deltaup y = h$. We define
\begin{align*}
  &x_i=x_\text{min}+i\,h, \qquad i=0,\ldots,N-1,
  \\&y_j=y_\text{min}+j\,h, \qquad j=0,\ldots,N-1,
\end{align*}
where $x_\text{min}=-L/2$ and $y_\text{min}=-L/2$ are offsets allowing us to centre the computation domain around the origin. We discretise time as $t_k=k\Deltaup t$, $k=0,1,2,\ldots$, so that $t_{k+1}=t_k+\Deltaup t$. Letting $u_{ij}^k\approx u(x_i,y_i,t_k)$ be the numerical approximation of the solution at position $(x_i,y_i)$ and time $t_k$, the numerical solution of Eq.~\eqref{reac-diff} is stepped in time by solving
\begin{align}\label{u-solve-discrete}
    u_{ij}^{k+1} = u_{ij}^k + \frac{D\Deltaup t}{h^2}\left(\left[u_{i-1,j}^k-2u_{ij}^k+u_{i+1,j}^k\right] + \left[u_{i,j-1}^k-2u_{ij}^k+u_{i,j+1}^k\right]\right) + \Deltaup t\,F(u_{ij}^k)
\end{align}
iteratively in $k$ for all indices $(i,j)$ for which $(x_i,y_j)\in\Omega$. Dirichlet boundary conditions are implemented by setting
\begin{align*}
    u_{ij}^k = u^* \quad \forall (i,j) \text{ such that $(x_i,y_j)\not\in\Omega$}, \quad \forall k.
\end{align*}
Neumann boundary conditions are only implemented in this work when $\Omega$ is a square domain, or a circular domain. For a square domain, Neumann boundary conditions are implemented using standard second order discretisation across the boundary, so that for points along the boundaries $i=0$, $i=N-1$, $j=0$, and $j=N-1$, Eq.~\eqref{u-solve-discrete} is used by defining
\begin{align*}
    u_{-1,j}^k \equiv u_{1,j}^k, \quad u_{N,j}^k \equiv u_{N-2,j}^k, \quad u_{i,-1}^k \equiv u_{i,1}^k, \quad u_{i,N}^k \equiv u_{i,N-2}^k.
\end{align*}
The initial condition is implemented by setting
\begin{align*}
  u_{ij}^0= \begin{cases} 0, &\quad \forall (i,j)\text{ such that $(x_i,y_i)\in\Omega$}
              \\u^*, &\quad \text{otherwise}.
            \end{cases}
\end{align*}
For Neumann boundary conditions, this represents the fact that $u(\b r, 0)=u^*\deltaup_{\p\Omega}(\b r)$ with $\int\d\b r\ u(\b r,0)=u^*|\partial\Omega|$.

\paragraph{Circular domain}
% See programs README.org under v10-1D
For circular domains $\Omega$ with radially symmetric solutions $u(r,t)$, where $r=|\b r|$, we implement a FTCS scheme by first expressing Eq.~\eqref{reac-diff} in polar coordinates and assuming circular symmetry. The reaction--diffusion equation becomes
\begin{align}\label{reac-diff-polar}
    u_t = D\frac{1}{r}(r u_r)_r + F(u) = D \left( u_{rr} + \frac{1}{r}u_r\right) + F(u).
\end{align}
It should be emphasised that this equation is well behaved as $r\to 0$. Indeed, by symmetry,
\begin{align}\label{polar-BC-origin}
    \lim_{r\to 0}u_r = 0,
\end{align}
so that using Bernoulli--L'Hôpital's rule, we have
\begin{align}\label{BH}
    \lim_{r\to 0} \frac{u_r}{r} = \lim_{r\to 0} u_{rr}.
\end{align}
To solve Eq.~\eqref{reac-diff-polar} numerically, we discretise the radial axis as $r_i=i\, h$, $i=0,\ldots N-1$, use second order finite difference for all spatial derivatives involved, and solve
\begin{align}\label{u-solve-discrete-polar}
  u_i^{k+1} = u_i^k + \begin{cases}
    \displaystyle\frac{2D\Deltaup t}{h^2}\left(u_{i-1}^k-2u_i^k+u_{i+1}^k\right) + \Deltaup t\,F(u_i^k), & i=0,
    \\\displaystyle\frac{D\Deltaup t}{h^2}\left(u_{i-1}^k-2u_i^k+u_{i+1}^k + \frac{1}{2i}\big[u_{i+1}^k-u_{i-1}^k\big]\right) + \Deltaup t\,F(u_i^k), & i=1,2,\ldots N-1,
          \end{cases}
\end{align}
iteratively in $k$. The time-stepping rule for $i=0$ is based on Eq.~\eqref{BH}, and $u_{-1}^k$ is set to $u_1^k$ to satisfy the Neumann-like symmetry condition~\eqref{polar-BC-origin} at the origin.

Dirichlet boundary conditions at the circular boundary $\p\Omega$ are implemented by setting $u_N^k=u^*$. Neumann boundary conditions are implemented by setting $u_{N}^k=u_{N-2}^k$ in Eq.~\eqref{u-solve-discrete-polar}.

\paragraph{Curvature and velocity}
Numerical estimates of the mean curvature $\kappa$ are calculated from:
\begin{align}\label{kappa}
    \kappa = \div\frac{\grad u}{|\grad u|} = \frac{u_{xx}{u_y}^2 - 2u_x u_y u_{xy} + u_{yy}{u_x}^2}{\left[{u_x}^2+{u_y}^2\right]^{3/2}},
\end{align}
using second-order-accurate centred finite differences for all first-order and second-order derivatives involved~\cite{sethian-1999}. Numerical estimates of the normal velocity are based on Eq.~\eqref{v-levelset}, which involves $\laplacian u$ and $|\grad u| = \sqrt{{u_x}^2+{u_y}^2}$. These derivatives are also estimated using second-order-accurate centred finite differences.

\paragraph{Domain boundaries}
Figure~\ref{fig:fronts} illustrates inward-moving fronts and outward-moving fronts on domains $\Omega$ specified by the following boundaries:
\begin{enumerate}
\item A circle of radius 1;
\item A square of side length 2;
\item A smooth cross shape parametrised in polar coordinates by the radial curve
    \begin{align*}
    R(\theta) = \sin^4(\theta) + \cos^4(\theta);
    \end{align*}
\item A petal with three branches (referred to below as 3-petal), parametrised in polar coordinates by the radial curve
\begin{align*}
    R(\theta) = R_0 + a\cos(n\theta), \qquad\text{where $R_0=0.65$, $a=0.35$, $n=3$}.
\end{align*}
\end{enumerate}
Inward-moving front simulations use a computational domain $[-1,1]\times[-1,1]$ ($L=2$). Outward-moving front simulations use a computational domain $[-8, 8]\times[-8,8]$ ($L=16$). Dirichlet boundary conditions are implemented for all the above domain shapes. Neumann boundary conditions are only implemented for the circular domain (in polar coordinates) and for the square domain (in Cartesian coordinates). Space and time units are arbitrary.

\section{Results and Discussion}
We start by examining numerical simulations of the two-dimensional analogue of the Fisher--KPP model with inward travelling fronts in the domains specified above. In this model, the reaction term represents logistic growth,
\begin{align}\label{fisher-kpp}
    F(u)=\lambda u(1-u).
\end{align}
In the infinite one-dimensional space, this model leads to the establishment of travelling wave solutions progressing with speed
\begin{align}\label{v1d-fisher}
    c = 2\sqrt{D\lambda},
\end{align}
provided the initial profile $u(x,0)$ has compact support~\cite{murray-1,edelstein-keshet,shigesada-kawasaki}.

\begin{figure}
        \includegraphics[width=\textwidth]{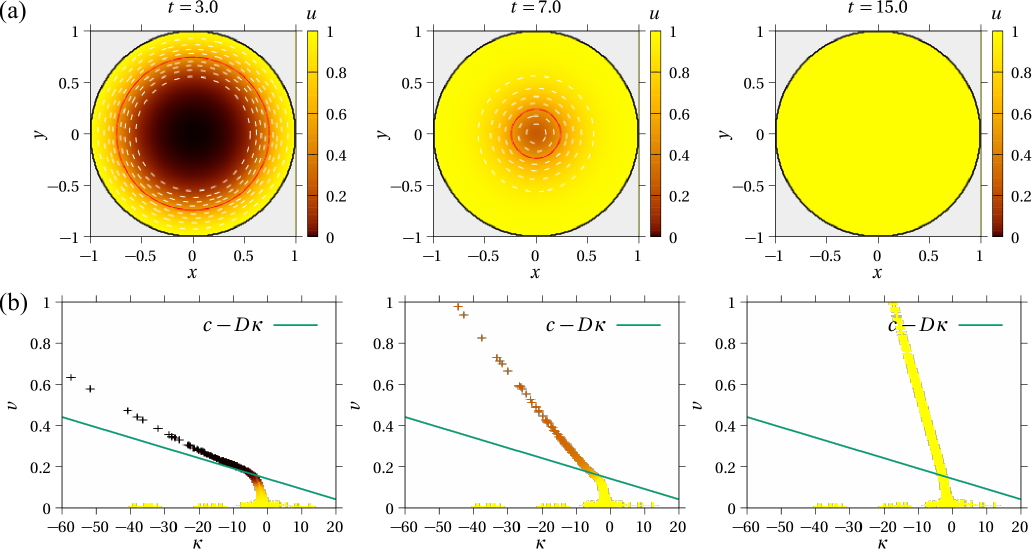}
        \caption{Curvature dependence of front speed in a circular pore ($F(u) = \lambda u(1-u)$, $\lambda=1$, $D=0.005$, $h=0.01$, $\Deltaup t=0.001$ and Dirichlet boundary conditions $u\vert_{\p\Omega}=1$). (a)~Snapshots of the solution $u(\b r,t)$ at different times. Contours $u=u_c$ are shown every 0.1 increment between $u_c=0$ and $u_c=1$ in dashed white lines, with $u_c=0.5$ shown in solid red. (b)~Front speeds $v(\kappa)$ determined numerically in the whole domain, coloured by the corresponding value of $u_c$ (pluses). The asymptotic expression $v(\kappa)\sim c-D\kappa$ from Eq.~\eqref{v-asymptotic} is shown as reference (green).}
        \label{fig:circle}
\end{figure}

Figure~\ref{fig:circle} shows snapshots at three different times of numerical simulations of Eq.~\eqref{reac-diff} with logistic source term~\eqref{fisher-kpp} on the circular domain and Dirichlet boundary conditions with $u^*=1$. The solution exhibits travelling fronts progressing inward into the circular pore space (Figure~\ref{fig:circle}a). Contour lines $u=u_c$ are shown every 0.1 increment of $u_c$. The contour line $u_c=0.5$, corresponding to locations with maximum growth rate, is shown in red.

Numerical estimates of front travelling speed and front curvature obtained at each (discretised) location $\b r$ of the domain are shown as data points $\big(v(\b r, t), \kappa(\b r, t)\big)$ in Figure~\ref{fig:circle}b. These data points are coloured by the value $u_c$ of the corresponding contour. At all times, there is a well defined linear correlation $v(\kappa)$ between front speed and curvature. At $t=3.0$, several data points with low density match the asymptotic relationship~\eqref{v-asymptotic}, shown in green, reasonably well. Numerical estimates of $\kappa$ and or $v$ at very low density at $t=3.0$ may be less accurate as they correspond to points near the centre of the domain, where spatial variations in $u$ are small. A value of $\kappa$ less than $-30$ correspond to radii of curvature less than about $3$ grid steps $h$. High density data points are close to the boundary and they deviate significantly from the relationship~\eqref{v-asymptotic}. At times $t=7.0$ and $t=15.0$, the relationship $v(\kappa)$ remains mostly linear, but the slope of the relationship is time-dependent, in contrast to Eq.~\eqref{v-asymptotic}.

\begin{figure}
        \includegraphics[width=\textwidth]{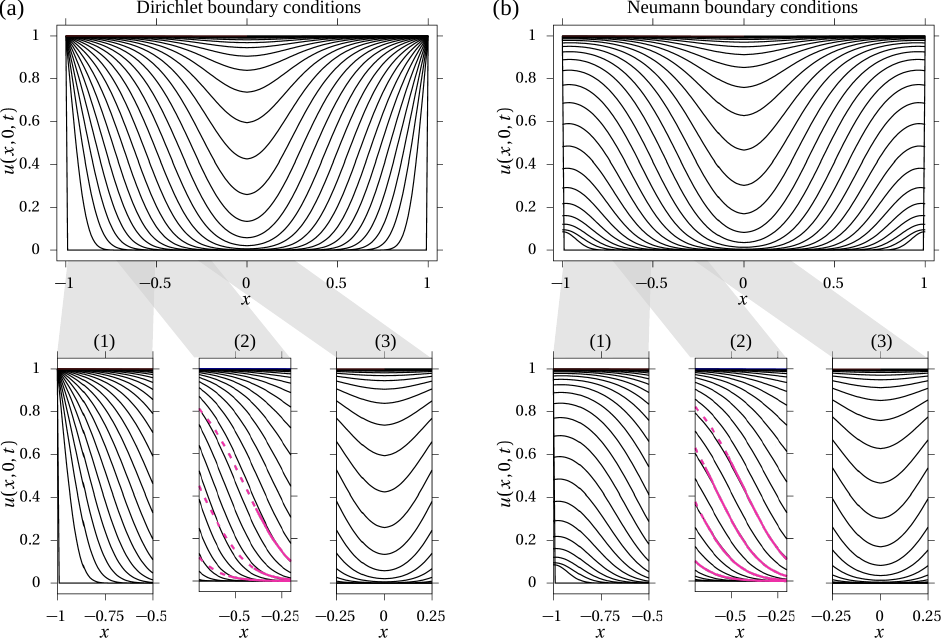}
        \caption{Comparison between solution profiles $u(x,0,t)$ in a $y=0$ cross-section of the circular domain (inward wave propagation) with either \cbeg (a) Dirichlet boundary conditions ($u\vert_{\p\Omega}=1$), or (b) Neumann boundary conditions\cend. Profiles are shown every 0.5 time units (black). Three phases are distinguished: (1) an establishment phase; (2) a travelling-wave-like phase, where solution profiles are similar to the one-dimensional travelling-wave solution profile (magenta), particularly at low values of $u$ (solid magenta line) ; (3) a dip-filling phase. The radially symmetric solutions $u(x,y,t)$ are computed numerically in polar coordinates on a domain $r\in [0,1]$. The one-dimensional travelling-wave solution profiles (magenta) are a mid-domain snapshot of the solution computed numerically in Cartesian coordinates on the domain $x\in [-8,8]$ with $u(-8,t)=1$, $u(8,t)=0$, suitably shifted along $x$.}
        \label{fig:phases}
\end{figure}

The results shown in Figure~\ref{fig:circle} suggest that the asymptotic growth law~\eqref{v-asymptotic} where the linear dependence of $v$ upon $\kappa$ has slope $-D$, and where $w$ in Eq.~\eqref{v} is given by the one-dimensional travelling wave speed $c$, Eq.~\eqref{v1d-fisher}, is well satisfied only during a limited time period, sufficiently far away from the domain boundary and before circular fronts collapse to a point. \cbeg Not all points of the domain may exhibit the behaviour~\eqref{v-asymptotic} at the same time, or at all. Equations~\eqref{vn}--\eqref{v1D} show that $w$ is well approximated by $c$ if the solution profile in the normal direction $\b n$ is close to the one-dimensional travelling-wave profile. To clarify which points of the domain satisfy this criterion in Figure~\ref{fig:circle}, we compare the solution profiles corresponding to Figure~\ref{fig:circle} with the one-dimensional travelling-wave profile in Figure~\ref{fig:phases}a. This comparison \cend confirms that the asymptotic growth law~\eqref{v-asymptotic} in Figure~\ref{fig:circle}b is well satisfied when and where the solution profile in the normal direction matches the one-dimensional travelling wave profile. Figure~\ref{fig:phases}b also shows numerical results of inward progressing fronts of the two-dimensional Fisher-KPP model obtained with Neumann boundary condition at $|\b r|=1$.

\paragraph{Wave propagation phases} From these simulations, we can identify three distinct phases of wave propagation during inward motion:
\begin{enumerate}
\item[(1)] An \emph{establishment phase}, where solution profiles are strongly influenced by the proximity of boundaries, and by the type of boundary condition implemented;
\item[(2)] A \emph{travelling-wave-like phase}, where solution profiles normal to the fronts closely match the travelling-wave profiles of the corresponding infinite, one-dimensional space solution;
\item[(3)] A \emph{dip-filling phase}, where fronts coming from opposite ends of the domain and travelling in opposite directions, collide and interact to build up the quantity $u$ symmetrically.
\end{enumerate}

\begin{figure}
        \includegraphics[width=\textwidth]{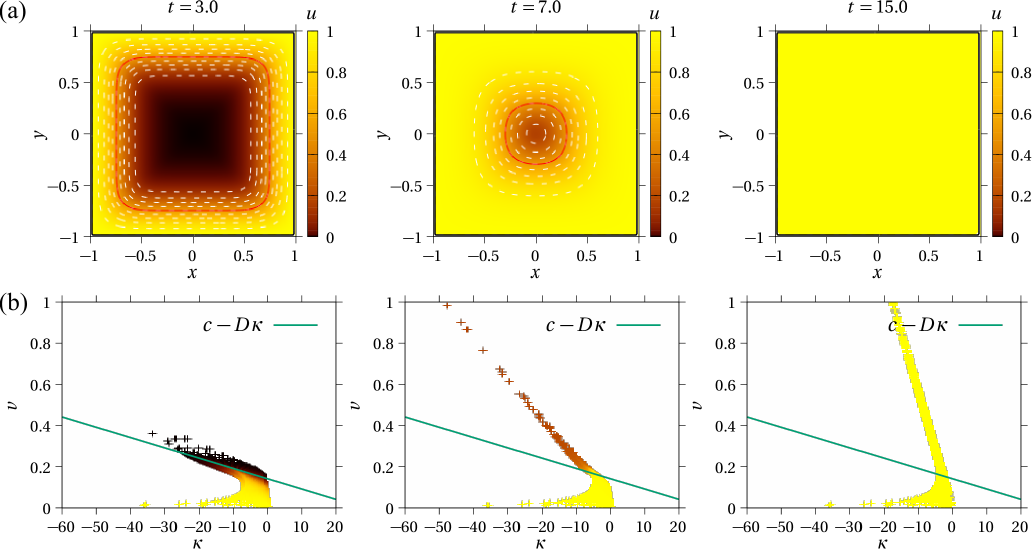}
        \caption{Curvature dependence of front speed in a square pore ($F(u) = \lambda u(1-u)$, $\lambda=1$, $D=0.005$, $h=0.01$, $\Deltaup t=0.001$ and Dirichlet boundary conditions $u\vert_{\p\Omega}=1$). Plot specifications are as in Figure~\ref{fig:circle}.}
        \label{fig:square}
\end{figure}

\begin{figure}
        \includegraphics[width=\textwidth]{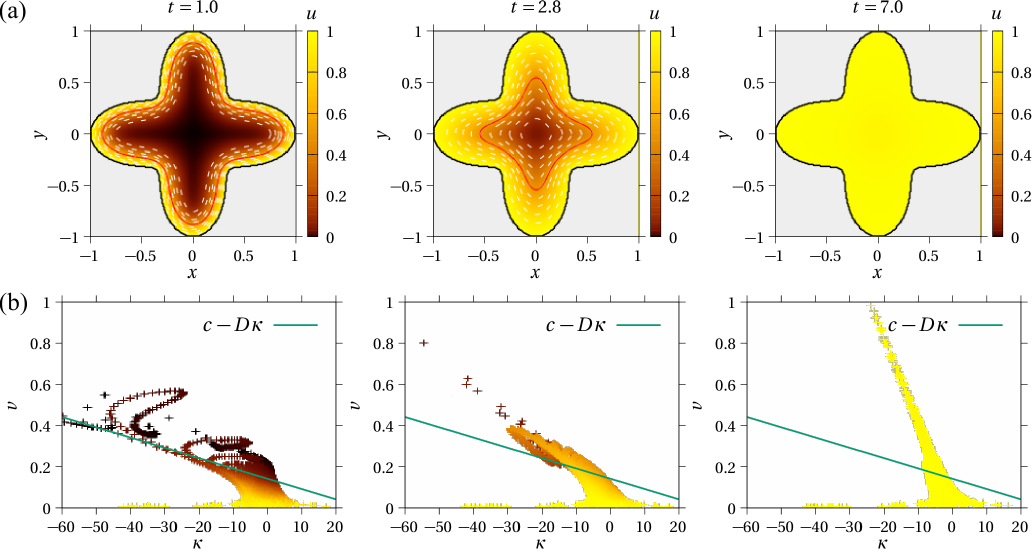}
        \caption{Curvature dependence of front speed in a cross-shaped pore with negative and positive curvature ($F(u) = \lambda u(1-u)$, $\lambda=1$, $D=0.005$, $h=0.01$, $\Deltaup t=0.001$ and Dirichlet boundary conditions $u\vert_{\p\Omega}=1$). Plot specifications are as in Figure~\ref{fig:circle}.}
        \label{fig:cross}
\end{figure}

\begin{figure}
        \includegraphics[width=\textwidth]{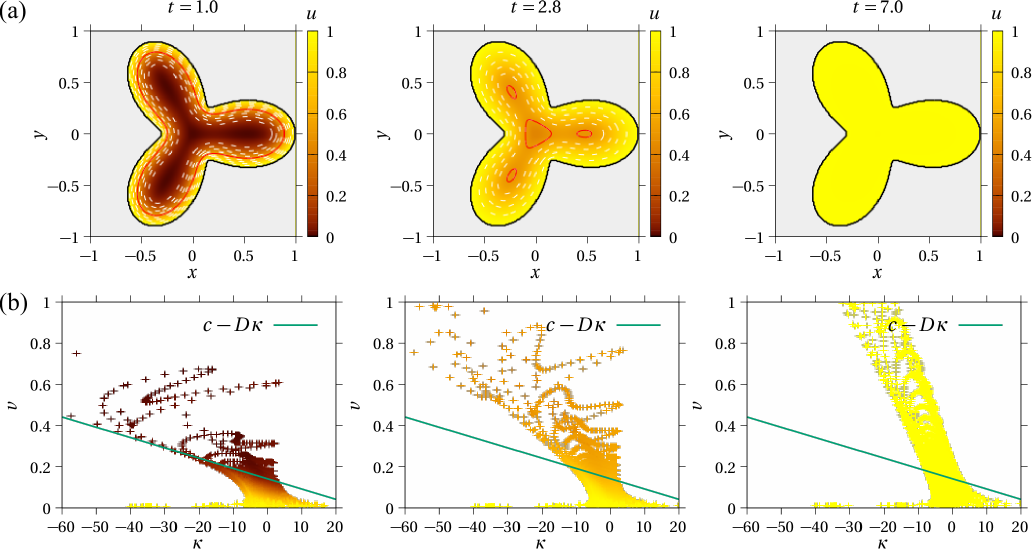}
        \caption{Curvature dependence of front speed in a 3-petal-shaped pore with negative and positive curvature ($F(u) = \lambda u(1-u)$, $\lambda=1$, $D=0.005$, $h=0.01$, $\Deltaup t=0.001$ and Dirichlet boundary conditions $u\vert_{\p\Omega}=1$). Plot specifications are as in Figure~\ref{fig:circle}.}
        \label{fig:petal3}
\end{figure}

\begin{figure}
        \centering\includegraphics[width=0.8\textwidth]{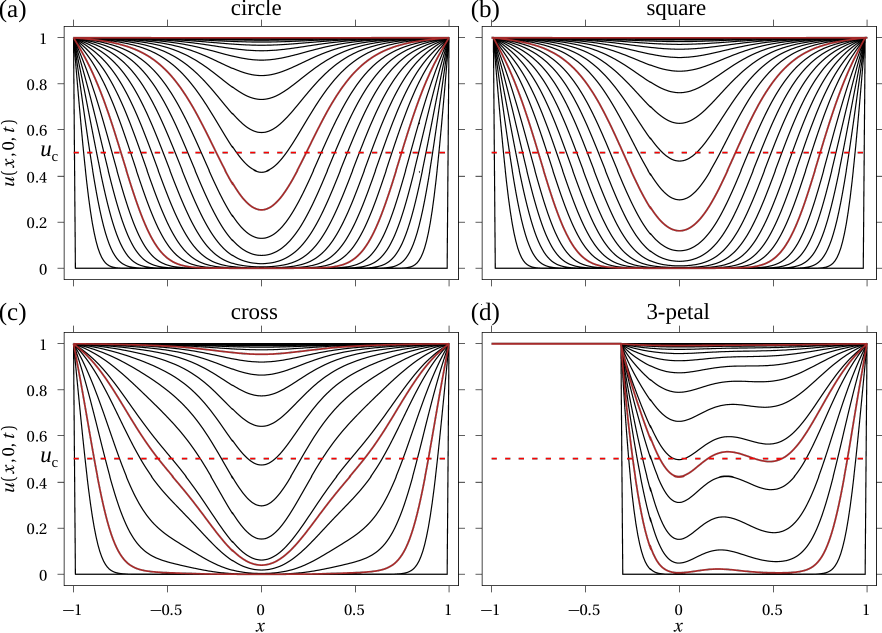}
        \caption{\cbeg Solution profiles $u(x,0,t)$ in a $y=0$ cross section of (a) the circular pore in Figure~\ref{fig:circle}; (b) the square pore in Figure~\ref{fig:square}; (c) the cross-shaped pore in Figure~\ref{fig:cross}; and (d) the 3-petal-shaped pore in Figure~\ref{fig:petal3}. \cend Profiles are shown every 0.5 time units (black) as well as at the times selected in Figures~\ref{fig:circle}--\ref{fig:petal3} (red). The intersection of the solution profiles with the $u_c=0.5$ level (red dashed line) corresponds to the travelling front shown as the red contour in Figures~\ref{fig:circle}--\ref{fig:petal3}.}
        \label{fig:cross-sections}
\end{figure}

Numerical simulations of inward progressing fronts of the Fisher--KPP model in the square pore shape, cross pore shape, and 3-petal pore shape with Dirichlet boundary conditions are shown in Figures~\ref{fig:square}, \ref{fig:cross}, and \ref{fig:petal3}, respectively. In all pores, travelling fronts tend to smooth their initial shape and to become circular toward the centre of the domain. This is consistent with the evolution of interfaces governed by mean curvature flow, whereby normal velocity depends linearly on curvature~\cite{brakke-1978,grayson-1987,giga-2006}. Figures~\ref{fig:square}b, \ref{fig:cross}b, and \ref{fig:petal3}b show that there is a clear correlation between $v$ and $\kappa$ that becomes increasingly linear as time progresses. However, as before, this relationship is strongly time dependent. The asymptotic relationship~\eqref{v-asymptotic} between $v$ and $\kappa$ is only well approximated during a limited period of time, away from the boundaries, and before fronts meet and interact in the centre. As for the circular pore shape, we can identify an establishment phase, a travelling-wave-like phase where solution profiles are similar to the one-dimensional travelling-wave profile (Figure~\ref{fig:cross-sections}) and where $v(\kappa)$ matches the asymptotic relationship~\eqref{v-asymptotic} well, and a dip-filling phase, characterised by a linear relationship $v(\kappa)$ but with a time-dependent slope. 

% At points of the domain close to the boundaries, front speed is not significantly influenced by front curvature, as can be seen in Figures~\ref{fig:circle}(b),\ref{fig:square}(b)--\ref{fig:petal3}(b) by data points with high value of~$u$ (yellow).

\cbeg The establishment phase results from complex interactions between boundary conditions, initial conditions, reaction, and diffusion. In biological invasions, the establishment phase represents the survival and growth of a small initial population comprising few individuals, before the population expands out in the surrounding environment. This phase typically depends on many environmental and demographic conditions~\cite{shigesada-kawasaki}. In our simulations, the initial population is distributed over the pore boundary. The establishment phase corresponds to the survival and growth of the density $u$ at every location of the boundary.

When a well-defined travelling-wave phase exists, transitions between the different phases may be estimated qualitatively from the width of the one-dimensional travelling wave front, i.e., the characteristic length over which the solution profile in the normal direction transitions from $u\approx 0$ to $u\approx 1$, and the one-dimensional travelling wave speed $c$. The establishment phase transitions into the travelling-wave phase when the solution profile approximates the shape of the one-dimensional travelling wave front over its characteristic width, and this solution profile detaches from the domain boundary. The travelling-wave phase transitions into the dip-filling phase when the width of the front begins to overlap with similar fronts travelling in opposite directions. In the Fisher-KPP model, the width of the travelling wave can be taken to be $4c$, the inverse of the steepness of the solution at $u=u_c$~\cite{murray-1}. This width is directly related to the travelling wave speed $c$.

\begin{figure} \centering\includegraphics[width=\textwidth]{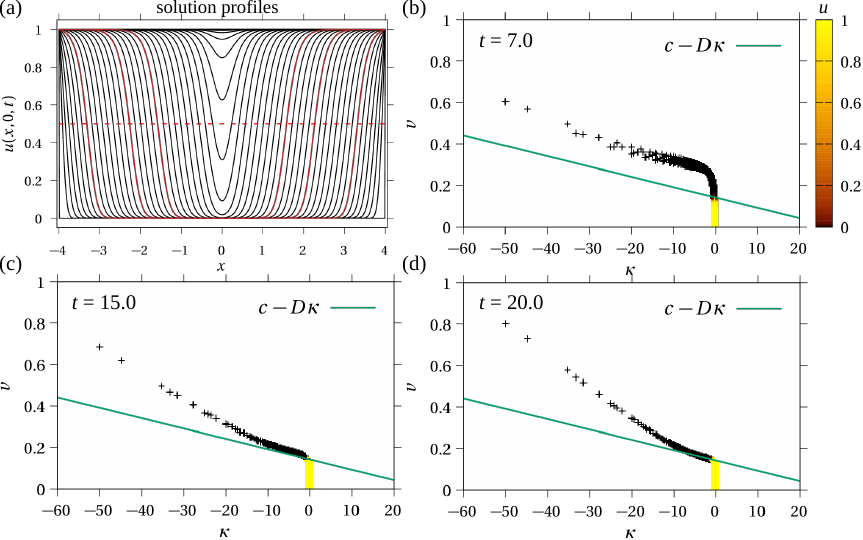}
        \caption{\cbeg Curvature dependence of front speed in a circular pore of radius~$4$ ($F(u) = \lambda u(1-u)$, $\lambda=1$, $D=0.005$, $h=0.01$, $\Deltaup t=0.001$, $u\vert_{\p\Omega}=1$). (a)~Solution profiles $u(x,0,t)$ are shown every 1 time unit (black) and at $t=7.0, 15.0, 20.0$ (red). (b)~Establishment phase: front speeds differ significantly from the asymptotic expression $v(\kappa)\sim c-D\kappa$ from Eq.~\eqref{v-asymptotic} (green). (c)--(d)~Travelling-wave phase: front speeds match the asymptotic expression $v(\kappa)\sim c-D\kappa$ from Eq.~\eqref{v-asymptotic} well, except at points far from the wave front, i.e., near boundaries of the domain where $u\approx 1$ (yellow pluses), and near the centre of the domain where curvature is high.\cend}
        \label{fig:circle-large}
\end{figure}

By decreasing diffusivity, or equivalently, by increasing domain size~\cite{buenzli-etal-2020}, the travelling-wave-like phase lasts longer, and the match between $v(\kappa)$ and the asymptotic relationship \eqref{v-asymptotic} improves (Figure~\ref{fig:circle-large}). In contrast, if diffusivity is too large or domain size too small, the solution may never exhibit a well-defined travelling-wave-like phase. In Figures~\ref{fig:circle} and~\ref{fig:circle-large}, the one-dimensional travelling wave front has the same width and speed. However, establishing a profile similar to that of the one-dimensional travelling wave takes longer in the larger circular pore, and results in a longer establishment phase (Figure~\ref{fig:circle-large}b).\cend

The normal velocity $v$ is not always a sole function of curvature. The spread of points $(\kappa, v)$ in Figures~\ref{fig:cross}b, \ref{fig:petal3}b indicates that other detail of the solution may influence the normal velocity. The curves formed by the spread of points in these figures are due to only calculating $v$ and $\kappa$ at discretised positions in the domain. In continuous space, the spread of points would fill a whole region, similar to what is seen at low curvature (more points of the computational grid have small curvature than high curvature).

\cbeg In Figure~\ref{fig:petal3}, the front $u=u_c$ splits into multiple disconnected fronts ($t=2.8$). Dip-filling occurs independently in the middle of the petals and in the centre of the domain. These separate dip-filling locations correspond to local maxima of the closest distance to the domain boundary~\cite{sethian-1999}.\cend

Outward-moving fronts clearly do not exhibit a dip-filling phase (Figure~\ref{fig:fronts}). The long-time solution $u(\b r, t)$ develops a profile in the normal direction that gradually converges to that of the one-dimensional infinite space solution~\cite{witelski-etal-2000,shigesada-kawasaki}. The asymptotic relationship~\eqref{v-asymptotic} is well satisfied sufficiently far from the boundaries, but the correction term due to curvature is small and tends to zero as $t\to\infty$. Indeed, as the solution expands out away from the origin, the curvature of the front decreases. In the following, we focus on analysing inward-moving fronts and the long-term behaviour of the solution in the dip-filling phase, since in this situation, the curvature of inward-moving fronts increases with time, and induces further dependences to the front speed, as suggested by our numerical simulations.

\paragraph{Normal diffusion in the dip-filling phase}
Figures~\ref{fig:circle}--\ref{fig:petal3} suggest that the dip-filling phase is similar across many domain shapes, and leads in the long term to a well-defined linear dependence $v(\kappa)$ of front speed upon curvature, although with a different slope compared to the asymptotic perturbation result~\eqref{v-asymptotic}. This similarity across domain shapes is due to the fact that the travelling fronts always evolve into shrinking circles in these examples, an evolution likely due initially to the explicit curvature-dependence of normal velocity in Eq.~\eqref{v}~\cite{brakke-1978,grayson-1987}. As time progresses, further curvature dependences of $v(\kappa)$ arise that reinforce this tendency.

\begin{figure}
        \centering\includegraphics[width=\textwidth]{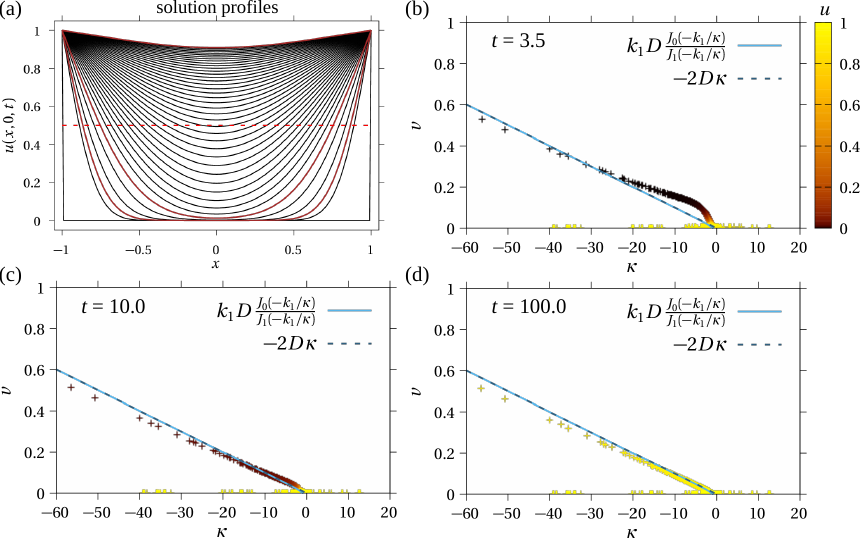}
        \caption{Curvature dependence of front speed in a circular pore for the diffusion equation with Dirichlet boundary  conditions ($F(u)=0$, $D=0.005$, $h=0.01$, $\Deltaup t=0.001$, $u\vert_{\p\Omega}=1$). \cbeg (a) Solution profiles are shown every 2.5 time units (black) as well at $t=3.5$, $t=10.0$, and $t=100.0$ (red). (b) \cend At early time ($t=3.5$) and small curvature (far from the centre), the relationship $v(\kappa)$ has a slope close to $-D$, which transitions into the asymptotic relationship $v(\kappa)\sim -2D\kappa$ at large curvature (see Eq.~\eqref{v-centre}). \cbeg (c)--(d) \cend With time, all fronts of the solution move with the time-independent speed $v(\kappa)=-2D\kappa$. }
        \label{fig:v-centre}
\end{figure}

To analyse these additional contributions to $v(\kappa)$ in the dip-filling phase, we now assume that the solution is approximately radially symmetric near the centre of the domain $r\to 0$. Since $\b n$ points in the opposite direction to the radial axis, $\p/\p n = \b n\vdot\grad = -\p/\p r$. The contribution to $v(\kappa)$ due to diffusion in the normal direction captured by the term $D u_{nn}/|\grad u|$ in Eq.~\eqref{vn} becomes, using Eqs~\eqref{polar-BC-origin}, and Bernoulli--L'Hôpital's rule~\eqref{BH}:
\begin{align*}
    D\frac{u_{nn}}{|\grad u|} \sim  D\frac{u_{rr}}{u_r} \sim D \frac{u_r/r}{u_r} \sim \frac{D}{r}, \qquad r\to 0.
\end{align*}
Since $\kappa=-1/r$, we obtain $D u_{nn}/|\grad u| \sim -D\kappa$ as $r\to 0$, so that the speed of travelling fronts in Eq.~\eqref{v} becomes
\begin{align}\label{v-centre}
    v(\kappa) \sim -2D \kappa - \frac{F(u)}{u_n}, \qquad r\to 0 \text{ or } \kappa\to -\infty.
\end{align}
We conclude that in the dip-filling phase, both the normal component of diffusion and the transverse component of diffusion contribute a term $-D\kappa$ each to front speed. While this is valid at all times for radially symmetric solutions as $\kappa\to -\infty$, it is only valid as $t\to\infty$ for non-circular domains since moving fronts in such domains evolve into circles only in the long-time limit. It is interesting to note here that some analyses of flame propagation also suggest flame displacement speed to be proportional to curvature with a factor twice the thermal diffusivity~\cite{Dave2020}. In Figure~\ref{fig:v-centre}, we illustrate this extra contribution of curvature to front speed in the dip-filling phase by presenting numerical simulations of the diffusion equation ($F(u)=0$) in a circular domain with Dirichlet boundary conditions $u^*=1$. These simulations show that without reaction terms, front speed in the dip-filling phase indeed converges to the time-independent linear relationship $v(\kappa)\sim -2D\kappa$. The speed of a travelling front, as it progresses in space, is time dependent due its changing curvature. The diffusion equation is an example for which the solution does not develop into a travelling wave in infinite space.

Importantly, the contribution to $v(\kappa)$ due to the normal component of diffusion is of the same order as the contribution due to the transverse component of diffusion obtained usually by singular perturbation analyses. These singular perturbations analyses thus do not describe the velocity of inward-moving circular fronts close to where they collide~\cite{tyson-keener-1988}; these developments are only valid away from boundaries and away from colliding fronts.

\paragraph{Reaction-rate dependence in the dip-filling phase} The contribution due to the reaction term in Eq.~\eqref{v-centre} also possesses a dominant linear dependence upon curvature. Indeed, proceeding similarly to above and using $u_r/r\sim u_{rr}$ as $r\to 0$ and $u_n = -u_r$, we have
\begin{align}
    \frac{1}{u_n(r,t)} \sim -\frac{1}{r u_{rr}(0,t)} = \frac{\kappa}{u_{nn}(0,t)}, \quad r\to 0\ \text{ or }\ \kappa\to-\infty.
\end{align}
Thus, in the dip-filling phase:
\begin{align}\label{v-dipfilling}
    v(\kappa) \sim -2D\kappa  - \frac{F(u\big(0,t)\big)}{u_{nn}(0,t)}\kappa, \qquad \kappa\to-\infty.
\end{align}
Equation~\eqref{v-dipfilling} shows that in the dip-filling phase, fronts evolve with a normal velocity dominantly proportional to curvature. The proportionality coefficient has an explicit, constant contribution due to diffusion, and a time-dependent contribution related to the autocatalytic evolution of the solution at the centre of the dip ($r=0$) governed by the reaction term $F$. It should be emphasised that the second term on the right hand side of Eq.~\eqref{v-dipfilling} may still contain implicit dependences upon $D$ via $u(0,t)$ and $u_{nn}(0,t)$.

\paragraph{Zero-diffusion limit}
Equations~\eqref{v-centre} and~\eqref{v-dipfilling} show that the dip-filling phase can be highly influenced by reaction terms since fronts of the solution may progress in space due to $u$ increasing locally (Figure~\ref{fig:conceptual}). However, the reaction-rate-dependent contribution to $v(\kappa)$ on the right hand side of Eq.~\eqref{v-dipfilling} is still coupled to diffusive effects. In this paragraph, we therefore investigate two remaining important points: (i) the influence of diffusion in the reaction-rate-dependent contribution to $v(\kappa)$; and (ii) how far from the origin $r\to 0$ we can expect the asymptotic behaviour of $v(\kappa)$ in Eq.~\eqref{v-dipfilling} to hold in practice.

To get insights into the velocity of fronts due to reaction terms only, we consider the zero diffusion limit $D\to 0$ in the dip-filling phase. We show that in this regime, the eikonal equation $v(\kappa)$ becomes time independent for any choice of $F(u)$. In other words, diffusive effects are entirely responsible for the time dependence of the slope of $v(\kappa)$ observed at long times in Figures~\ref{fig:circle},\ref{fig:square}--\ref{fig:petal3}, and captured in Eq.~\eqref{v-dipfilling}. First, let us show that the velocity field $v(\b r, t)$ of all moving fronts is independent of time if $D=0$. From Eq.~\eqref{v-levelset} with $D=0$, we have
\begin{align}\label{vt0-diffzero}
    v_t(\b r, t) = \frac{F'(u)u_t}{|\grad u|} - \frac{F(u)}{|\grad u|^2}(|\grad u|)_t = \frac{F'(u)F(u)}{|\grad u|} - \frac{F(u)}{|\grad u|^2}F'(u)|\grad u| = 0,
\end{align}
where we have used
\begin{align*}
    |\grad u|_t = \pd{}{t}\sqrt{\grad u\vdot\grad u} = \frac{\grad u_t\vdot\grad u}{|\grad u|} = \frac{F'(u)\grad u\vdot \grad u}{|\grad u|} = F'(u)|\grad u|.
\end{align*}
In the dip-filling phase where travelling fronts become circular, the curvature of a front of radius $r$ is $-1/r$. The curvature of this front increases with time (in absolute value) as the inward-moving front comes closer to the origin $r=0$. At a fixed location in space, however, the curvature map $\kappa(\b r, t)$ for circular fronts about the origin is simply given by $\kappa(\b r, t) = -1/r$, where $r=|\b r|$, and it is therefore independent of time. Since both the velocity map $v(\b r, t)$ and curvature map $\kappa(\b r, t)$ are time independent in the zero-diffusion limit of the dip-filling phase, the relationship $v(\kappa)$ is also time independent. We can conclude that the steepening with time of the linear relationship $v(\kappa)$ in our numerical simulations in Figures~\ref{fig:circle}--\ref{fig:petal3} is entirely due to diffusive effects.

\begin{figure}
    \centering\includegraphics[width=\textwidth]{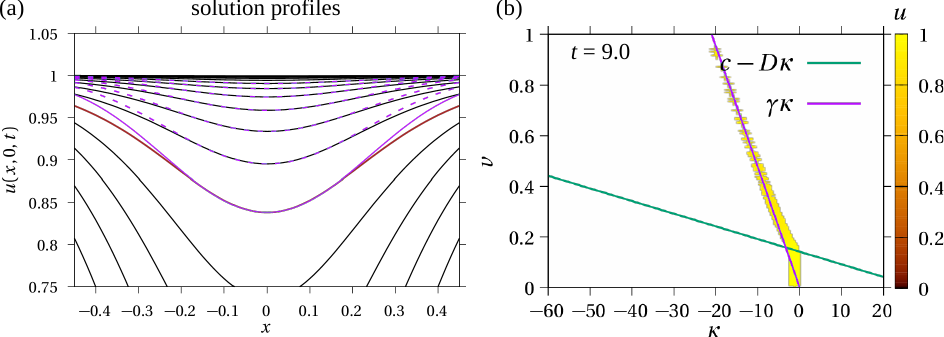}
    \caption{Curvature dependence of front speed when diffusion is turned off at $t=9.0$ in the two-dimensional analogue of the Fisher--KPP model with Dirichlet boundary conditions ($F(u)=\lambda u(1-u)$, $\lambda=1$, $D=0.005$, $h=0.01$, $\Deltaup t=0.001$, $u\vert_{\p\Omega}=1$). (a) Solution profiles are shown every 0.5 time units (black) and at $t=9.0$ (red). The parameter $\gamma$ in Eq.~\eqref{u-diffzero} that best matches the solution profile at $t=9.0$ near the domain centre (solid \cbeg magenta \cend curve) predicts the slope of the relationship $v(\kappa)=-\gamma\kappa$ very well (\cbeg magenta \cend line in (b)). Dashed \cbeg magenta \cend lines show $U(r,t)$ given by Eq.~\eqref{u-diffzero} every 0.5 time units from $t=9.5$ to $t=12.0$. The constant $A_0\approx 0.838$ in Eq.~\eqref{u-diffzero} is taken from the numerical solution $u(0,9.0)$. For increased accuracy in (b), curvature $\kappa$ at the radial coordinate $r$ is taken to be $-1/r$ rather than being estimated numerically from the solution $u$.}
    \label{fig:dipfilling-diffzero}
\end{figure}

Figure~\ref{fig:dipfilling-diffzero} shows numerical simulations in which the two-dimensional Fisher--KPP model is evolved in a circular pore until time $t=9.0$, at which point diffusion is set to zero. From this point onwards, the solution continues to evolve due to logistic growth only~(Figure~\ref{fig:dipfilling-diffzero}a). We observe that the relationship $v(\kappa)$ in the dip-filling phase is linear with a slope that becomes steeper with time until $t=9.0$. After this point, the relationship $v(\kappa)$ no longer evolves in time and is approximately given by $v(\kappa) = -\gamma \kappa$, where $\gamma>0$ is a constant (Figure~\ref{fig:dipfilling-diffzero}b).

The proportionality constant $\gamma$ is directly related to the profile of the solution at $t=9.0$ and its subsequent evolution at times $t>9.0$, as follows. Let $t_0$ be the time at which $D$ is set to zero. When $D=0$ and with radial symmetry (dip-filling phase), the solution profile at time $t>t_0$ is given by a function $U(r,t)$, where $r\geq 0$ is the radial coordinate. From Eq.~\eqref{v-levelset} and assuming $v(\kappa) = -\gamma\kappa = \gamma /r$, we have:
\begin{align*}
  v = \frac{F(U)}{U_r} = \frac{\gamma}{r},
\end{align*}
or, equivalently,
\begin{align}\label{u-profile}
    \frac{U_r}{F(U)} = \frac{r}{\gamma}.  
\end{align}
Integrating the differential equation in $r$~\eqref{u-profile} for $U(r,t)$ can be done explicitly for several choices of the reaction term $F(u)$. For logistic growth $F(u)=\lambda\,u(1-u)$, $1/F(u) = \lambda^{-1}(\d/\d u)\left[\ln(u)-\ln(1-u)\right]$, so that
\begin{align}
  U(r,t) = \frac{1}{1+A(t)\exp{-\frac{\lambda}{2\gamma}r^2}},
\end{align}
where the integration constant $A(t)=1/U(0,t)-1$ contains the only time dependence. Differentiating this solution with respect to $t$ gives
\begin{align*}
    U_t = -\frac{A_t}{A}U(1-U),
\end{align*}
which shows that $A(t) = A_0\exp\{-\lambda(t-t_0)\}$. Therefore, for $t>t_0$,
\begin{align}\label{u-diffzero}
    U(r,t) = \frac{1}{1+A_0\exp\left\{-\frac{\lambda}{2\gamma}r^2-\lambda(t-t_0)\right\}},
\end{align}
which means that the solution converges exponentially fast to the uniform asymptotic solution $\lim_{t\to\infty}U(r,t) = 1$, with contour lines that move inward with velocity $v(r)=\gamma/r$.

The value of $\gamma$ that best fits the numerical solution profile at $t=9.0$ around $r=0$ in Figure~\ref{fig:dipfilling-diffzero}a, is $\gamma\approx 0.047$. This value of $\gamma$ found from the solution profile predicts the slope of the growth law $v(\kappa)$ very well, see Figure~\ref{fig:dipfilling-diffzero}b. The solution in Eq.~\eqref{u-diffzero} differs from the numerical solution at locations where fronts have small curvature ($|x|\gtrsim 0.2$, corresponding to $\kappa \lesssim 5$, see Figure~\ref{fig:dipfilling-diffzero}a). At these curvatures, numerical estimates of $v(\kappa)$ in Figure~\eqref{fig:dipfilling-diffzero}b deviate from the linear relationship $v(\kappa)=-\gamma\kappa$. Figure~\ref{fig:dipfilling-diffzero}a also shows that once diffusion is turned off, the time evolution of the solution is very well represented by Eq.~\eqref{u-diffzero}.

The correspondence in Eq.~\eqref{u-profile} between the eikonal equation $v(\kappa)$ observed at a fixed time and solution profiles in the radial direction around the origin may be generalised to time-dependent and nonlinear relationships $v(\kappa)$ by integrating, from Eq.~\eqref{v},
\begin{align*}
      \frac{U_r}{F(U)} = \frac{1}{v(-1/r) - 2D/r},
\end{align*}
with respect to $r$, for example using numerical quadrature. This correspondence may be useful to predict the speed of travelling fronts in the dip-filling phase from a single snapshot in time of the solution profile, such as that provided by a biological experiment. On the other hand, measuring curvature-dependent speeds of travelling fronts may allow the estimation of solution profiles that may otherwise be difficult to measure, such as in wildfires. 

In summary, the dip-filling phase is characterised by travelling front velocities $v(\kappa)$ that are affected by curvature due to several contributions: (i) the transverse component of diffusion, always exactly contributing a term $-D\kappa$; (ii) the normal component of diffusion, contributing an extra term $-D\kappa$ at large curvature $\kappa\to -\infty$; (iii) reaction terms, which provide additional contributions to the normal speed that also becomes linear in $\kappa$ at large curvature $\kappa\to -\infty$, with a slope whose time-dependence is entirely due to diffusive effects.

\paragraph{Linearised models}
To understand in more detail how solution profiles converge in the dip-filling phase to profiles that precisely give rise to linear contributions in the growth law $v(\kappa)$, and to exhibit explicit time dependences of the slope of these linear contributions in the long-time limit, we now consider two relevant linear models: Skellam's model, where $F(u)=\lambda u$, and a saturated-growth model, where $F(u)=\lambda(1-u)$. Skellam's model corresponds to linearising the logistic reaction rate about the unstable rest state $u=0$, and is thus expected to describe the two-dimensional analogue of the Fisher-KPP model at early times. The saturated growth model corresponds to linearising the logistic reaction rate about the stable excited state $u=1$ reached by the solution of the Fisher-KPP model in the long-time limit.

Since our focus is the dip-filling phase, where solutions develop circular fronts in the long term, we assume radially symmetric solutions and solve Eq.~\eqref{reac-diff} for a radially symmetric solution $u(r,t)$ with
\begin{align}\label{skellam}
    F(u)=\lambda u.
\end{align}
The solution can be found by using the Fourier-Laplace transform of $u(r,t)$ in time, and solving the resulting eigenvalue problem of the Laplacian in polar coordinates with radial symmetry. The general solution is a superposition of Bessel functions of the first and second kind~\cite{abramowitz-stegun}. Only Bessel functions of the first kind, $J_n(x)$, are well defined at $x=0$, and only the angular mode $n=0$ is radially symmetric, so solutions have the form
\begin{align*}
    u(r, t) = \left[a_0 + \sum_{s=1}^\infty a_{k_s} J_0(k_sr)\exp\{-k_s^2Dt\}\right]\exp\{\lambda t\},
\end{align*}
where $a_0$ and $a_{k_s}$, $s=1,2,\ldots$, are constants that depend on the initial profile $u(r,0)$. The radial modes $k_s$ are determined by the boundary condition at $r=R$. If $\lambda=0$, the expression above is the general solution of the diffusion equation with radial symmetry. In this case, we can satisfy the Dirichlet boundary condition $u(R,t)=u^*=1$ by setting $a_0=1$ and $k_s$ such that $k_s R = j_{0,s}$, where $j_{0,s}>0$, $s=1,2,\ldots$, are the roots of the Bessel function $J_0$. If $\lambda >0$, it is more consistent to satisfy a time-dependent Dirichlet boundary condition $u(R,t)=\exp\{\lambda t\}$, in which case $a_0=1$ and $k_s$ are as above. Alternatively, given that $J_0'(x)=-J_1(x)$, we can enforce Neumann boundary conditions by selecting modes $k_s'$ such that $k_s' R = j_{1,s}$, where $j_{1,s}>0$, $s=1,2,\ldots$, are the roots of the Bessel function $J_1$.

To solve the equivalent problem with saturated growth reaction term
\begin{align*}
    F(u)=\lambda(1-u),
\end{align*}
we let $w=1-u$, so that $w_t = D\laplacian w - \lambda w$ is Skellam's model with the opposite sign for $\lambda$. The solution $u=1-w$ with Dirichlet boundary conditions is thus given by
\begin{align*}
      u(r, t) = 1 - \sum_{s=1}^\infty a_{k_s}J_0(k_s r)\exp\{-k_s^2D t\}\exp\{-\lambda t\}.
\end{align*}
At leading order in the long-time limit $t\to\infty$, we thus have, for the diffusion model ($u_\text{diff}$), Skellam's model with Dirichlet ($u_\text{Skellam}^\text{D}$) and Neumann ($u_\text{Skellam}^\text{N}$) boundary conditions, and for the saturated growth model ($u_\text{sat}$):
\begin{align}
    u_\text{diff}(r, t) &\sim 1 - a_{k_1} J_0(k_1r)\exp\{-k_1^2Dt\}, &&(F(u)=0, \  u(1,t)=1),\label{u-longtime-diff}
  \\u_\text{Skellam}^\text{D}(r, t) &\sim \left[1 - a_{k_1} J_0(k_1r)\exp\{-k_1^2Dt\}\right]\exp\{\lambda t\}, &&(F(u)=\lambda u, \  u(1,t)=\exp\{\lambda t\}),\label{u-longtime-skellam-dirichlet}
    \\u_\text{Skellam}^\text{N}(r, t) &\sim \left[a_0 - a_{k_1'} J_0(k_1'r)\exp\{-k_1'^2Dt\}\right]\exp\{\lambda t\}, &&(F(u)=\lambda u, \  u_r(1,t)=0),\label{u-longtime-skellam-neumann}
    \\u_\text{sat}(r, t) &\sim 1 - a_{k_1}J_0(k_1 r)\exp\{-k_1^2Dt\}\exp\{-\lambda t\}, &&(F(u)=\lambda(1-u), \  u(1,t)=1),\label{u-longtime-saturatedgrowth}
\end{align}
where the signs in front of $a_0, a_{k_1}$ and $a_{k_1'}$ are set such that these constants are all positive, and where we assume $R=1$, so that $k_1=j_{0,1}\approx 2.405$, and $k_1'=j_{1,1}\approx 3.832$. We can use the explicit long-term solutions in Eq.~\eqref{u-longtime-diff}--\eqref{u-longtime-saturatedgrowth} to estimate the speed of moving front in the dip-filling phase according to Eq.~\eqref{v}--\eqref{vn}. Using $\p/\p n = -\p/\p r$, we have
\begin{align*}
    - D\frac{u_{nn}}{u_n} = D\frac{u_{rr}}{u_r} \sim k_1D \frac{J_0''(k_1r)}{J_0'(k_1r)}, \qquad t\to\infty,
\end{align*}
(with $k_1$ replaced by $k_1'$ for Skellam's model with Neumann boundary conditions). From Bessel's equation satisfied by $J_0$, $J_0''(x)/J_0'(x) + 1/x + J_0(x)/J_0'(x) = 0$, so that substituting $x=k_1 r$ and using $\kappa = -1/r$ and $J_0'(x)=-J_1(x)$~\cite{abramowitz-stegun}, we obtain
\begin{align*}
    - D\frac{u_{nn}}{u_n} \sim  D\kappa + k_1D \frac{J_0(-k_1/\kappa)}{J_1(-k_1/\kappa)}, \qquad t\to\infty,
\end{align*}
in all four models. The velocity of fronts is thus given by
\begin{align}
  v(\kappa) &\sim k_1D \frac{J_0(-k_1/\kappa)}{J_1(-k_1/\kappa)} + \frac{F(u)}{u_r}, &&t\to\infty,&\label{v-linear-models}
  \\&\sim -2D\kappa + \frac{F(u)}{u_r}, &&t\to\infty, \quad\kappa\to-\infty,&\notag
\end{align}
where the last asymptotic expression uses $J_0(x) = 1-x^2/4 + \Order(x^4)$, and $J_1(x)=x/2 + \Order(x^3)$ as $x\to 0$~\cite{abramowitz-stegun}. This result is an explicit verification of the asymptotic behaviour derived in Eq.~\eqref{v-centre}. Eq.~\eqref{v-linear-models} also provides the contribution to $v(\kappa)$ due to diffusion for any curvature in the long-time limit, but the difference with the large-curvature expression is very small (Figure~\ref{fig:v-centre}).

We can further elucidate the contribution to $v(\kappa)$ due to the reaction term, $F(u)/u_r$, in the long-time limit based on Eqs~\eqref{u-longtime-diff}--\eqref{u-longtime-saturatedgrowth}. In Skellam's model with Dirichlet boundary condition, we obtain
\begin{align*}
  v_\text{Skellam}^D(\kappa) &\sim \Big(k_1D-\frac{\lambda}{k_1}\Big)\frac{J_0(-k_1/\kappa)}{J_1(-k_1/\kappa)} + \frac{\lambda\exp\{k_1^2Dt\}}{k_1a_{k_1}J_1(-k_1/\kappa)}, &&t\to\infty,&
    \\&\sim -2D\kappa - \frac{2\lambda}{k_1^2}\left(\frac{\exp\{k_1^2D t\}}{a_{k_1}}-1\right)\kappa, &&t\to\infty, \kappa\to -\infty.
\end{align*}

In the saturated growth model, by retaining the next-order term in the long-time limit, we obtain:
\begin{align*}
  v_\text{sat}(\kappa) &\sim \left(k_1D+\frac{\lambda}{k_1}\left[1+\mathcal{J}(\kappa)\e^{\displaystyle-(k_2^2\!-\!k_1^2)Dt}\right]\right)\frac{J_0(-k_1/\kappa)}{J_1(-k_1/\kappa)}, &&t\to\infty,&
    \\&\sim -2D\kappa - \frac{2\lambda}{k_1^2}\left(1+\frac{a_{k_2}}{a_{k_1}}\left(1-\frac{k_2^2}{k_1^2}\right)\e^{\displaystyle-(k_2^2\!-\!k_1^2)Dt}\right)\kappa, &&t\to\infty, \kappa\to -\infty.
\end{align*}
where 
\begin{align*}
  \mathcal{J}(\kappa) &= \frac{a_{k_2}}{a_{k_1}}\left(\frac{J_0(-k_2/\kappa)}{J_0(-k_1/\kappa)} - \frac{k_2J_1(-k_2/\kappa)}{k_1 J_1(-k_1 /\kappa)}\right)
  \\&\sim \frac{a_{k_2}}{a_{k_1}}\left(1-\frac{k_2^2}{k_1^2}\right), \qquad\kappa\to-\infty.
\end{align*}
It can also be checked explicitly from Eqs~\eqref{u-longtime-diff}--\eqref{u-longtime-saturatedgrowth} that the reaction-dependent expressions found above by calculating $F(u)/u_r$ in the limit $r\to 0$, correspond exactly to $-\left[F\big(u(0,t)\big)/u_{rr}(0,t)\right]\kappa$, as predicted by Eq.~\eqref{v-dipfilling}.

\begin{figure}
        \centering\includegraphics[width=\textwidth]{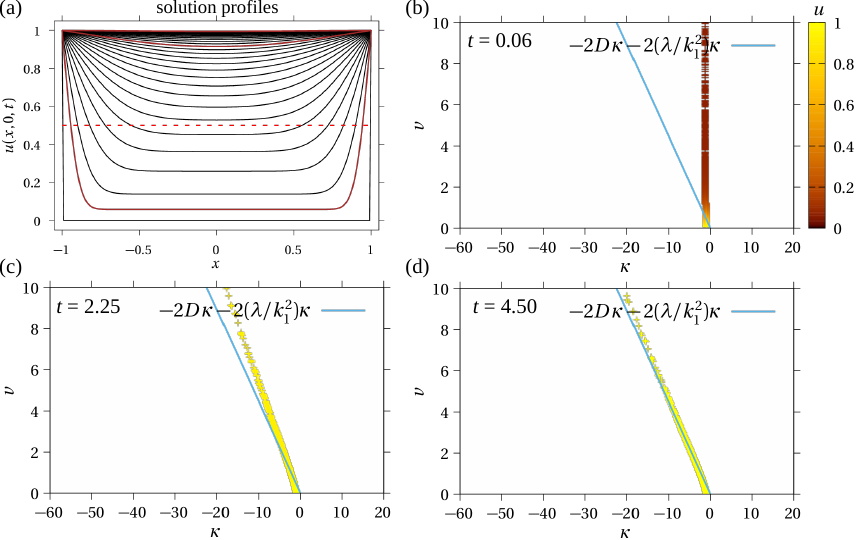}
        \caption{Curvature dependence of front speed in a circular pore for the saturated growth model with Dirichlet boundary conditions ($F(u)=\lambda(1-u)$, $\lambda=1$, $D=0.05$, $h=0.01$, $\Deltaup t=0.0001$,  $u\vert_{\p\Omega}=1$). \cbeg (a)\cend~Solution profiles are shown every 0.15 time units (black) as well as at the times $t=0.06, 2.25, 4.50$ (red). \cbeg (b)--(d)~With time, all fronts of the solution move with a speed approaching the asymptotic limit $v_\text{Sat}(\kappa)$ in Eq.~\eqref{v-satgrowth} (blue). \cend In these simulations, diffusivity is increased by a factor 10 compared to previous inward moving front simulations, to prevent the reaction term from dominating the evolution in the dip-filling phase.}\label{fig:saturated-growth}
\end{figure}

\begin{figure}
        \centering\includegraphics{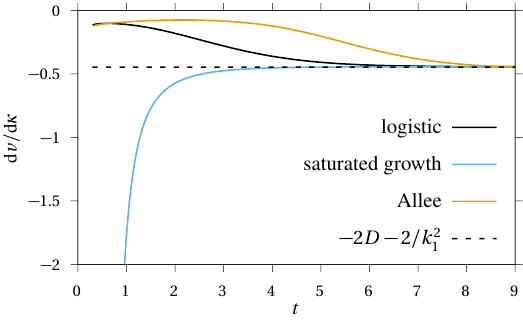}
        \caption{Slope of the linear relationship $v(\kappa)$ in the dip-filling phase versus time. Both the two-dimensional analogue of the Fisher-KPP model and the saturated growth model converge to the linear relationship $v(\kappa)$ given by Eq.~\eqref{v-satgrowth}. In these simulations all models are solved on the circular domain with Dirichlet boundary conditions and $D=0.05$, $h=0.05$, $\Deltaup t = 0.001$.}\label{fig:slope-vs-t}
\end{figure}

We see that in Skellam's model, the slope of the linear relationship $v(\kappa)$ continues to increase with time (in absolute value). In contrast, in the saturated growth model, the slope of the linear relationship $v(\kappa)$ decreases with time (in absolute value), and converges to
\begin{align}\label{v-satgrowth}
    \lim_{t\to\infty} v_\text{Sat}(\kappa) = -2D\kappa - \frac{2\lambda}{k_1^2}\kappa,
\end{align}
as confirmed by our numerical simulations (Figures~\ref{fig:saturated-growth}, \ref{fig:slope-vs-t}).
Figure~\ref{fig:slope-vs-t} shows that the front speed law $v(\kappa)$ in the two-dimensional Fisher-KPP model also converges to Eq.~\eqref{v-satgrowth} in the long-time limit, which is consistent with the fact that the saturated growth model is the linearisation of the Fisher-KPP model about the steady state $u=1$. At early times, however, the Fisher-KPP model is similar to Skellam's model. The slope of $v(\kappa)$ therefore first increases in time, before converging to that of Eq.~\eqref{v-satgrowth} in all pore shapes (Figures~\ref{fig:circle},\ref{fig:square}--\ref{fig:petal3}).

\begin{figure}
        \centering\includegraphics[width=\textwidth]{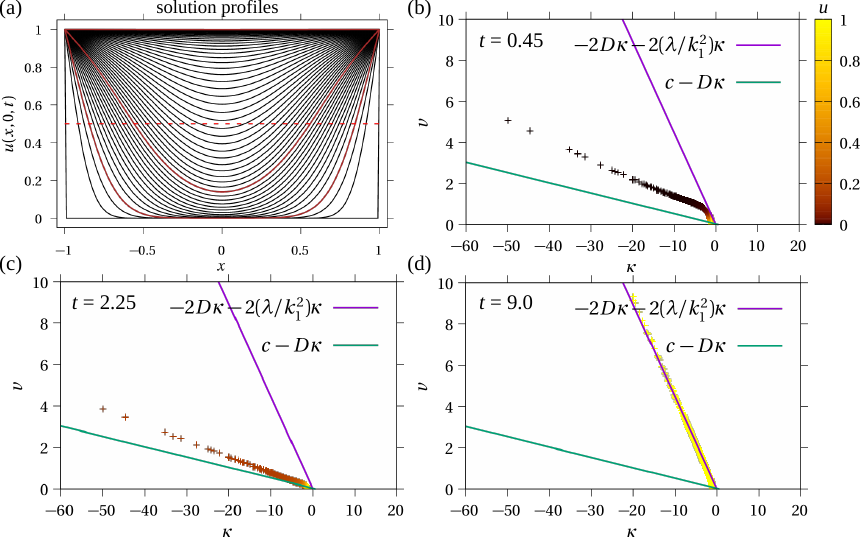}
        \caption{Curvature dependence of front speed in a circular pore for the strong Allee growth model with Dirichlet boundary conditions ($F(u)=\lambda u(u-a)(1-u)$ with $a=2/5$, $\lambda=1/(1-a)=3/5$, $D=0.05$, $h=0.01$, $\Deltaup t = 0.0001$, $u\vert_{\p\Omega}=1$). For this model, the minimum speed of travelling waves in infinite onedimensional space is $c=\sqrt{2D\lambda}(1/2-a)$. \cbeg (a)\cend~Solution profiles are shown every 0.15 time units (black) as well as at the times $t=0.45, 2.25, 9.0$ (red). \cbeg (b)--(d)~With time, all fronts of the solution move with a speed approaching the asymptotic limit $v_\text{Sat}(\kappa)$ in Eq.~\eqref{v-satgrowth} (magenta). The asymptotic expression $v(\kappa)\sim c-D\kappa$ from Eq.~\eqref{v-asymptotic} is also shown for reference (green). \cend Diffusivity is the same as in the saturated growth model for comparison in Figure~\ref{fig:slope-vs-t}.}\label{fig:Allee}
\end{figure}

\paragraph{Strong Allee effect} Finally, we consider an example of bistable reaction term where both the rest state $u=0$ and excited state $u=1$ are stable steady states. A classic example is provided by population growth with strong Allee effect, modelled by $F(u)=\lambda u(u-a)(1-u)$ with $0<a<1$. This model supports travelling waves with speed
\begin{align*}
  c=\sqrt{2D\lambda}\left(\frac{1}{2}-a\right)
\end{align*}
in one-dimensional infinite space under appropriate boundary and initial conditions~\cite{lewis-kareiva-1993}. In higher dimensions, the invasion of available space by the solution depends on the size and on the shape of the initial condition~\cite{lewis-kareiva-1993,Li2022}. Numerical simulations of this model are shown in Figure~\ref{fig:Allee} with $a=2/5$ and $\lambda=1/(1-a)=5/3$ so that $F(u)\sim 1-u$ as $u\to 1$ and $c>0$. Here too, the relationship $v(\kappa)$ converges to that of the saturated growth model in Eq.~\eqref{v-satgrowth} as $t\to \infty$.

Our numerical observation that the linear relationship $v(\kappa)$ of the two-dimensional analogue of the Fisher-KPP model and of the strong Allee effect converge to that of the saturated growth model in the long time is not trivial, even if the reaction terms in these models becomes similar in this limit. From Eq.~\eqref{v-dipfilling}, the slope of the relationship $v(\kappa)$ depends on $u_{rr}(0,t)$. This quantity could in principle depend on how the solution profile approaches steady-state from the initial condition, where the reaction term has different contributions in the three models.

\section{Conclusions \cbeg and future work \cend}
Reaction--diffusion waves propagating in excitable systems in multiple spatial dimensions are strongly influenced by travelling front curvature. The eikonal equation for these parabolic systems of differential equations is an expression that determines the local normal velocity of travelling fronts. In principle, a complete description of the eikonal equation at each location of the domain and each time enables one to fully evolve the solution in time, much like the method of characteristics for hyperbolic conservation laws~\cite{lax-1973}. The eikonal equation thus provides valuable physical insight into the behaviour of reaction--diffusion systems, as well as mathematical insights from known results from geometric flows that evolve moving interfaces based on velocity laws~\cite{grayson-1987,andrews-etal-2020}. In a tissue engineering context, the eikonal equation provides a biological growth law that describes how bulk tissue production and diffusive redistribution of tissue material lead to differential progression rates of tissue interfaces during tissue growth or invasion~\cite{goriely-2017,buenzli-etal-2020,browning-etal-2021}.

In the present work, we analysed the eikonal equation for a single reaction--diffusion equation with arbitrary reaction term. We showed that the contribution $-D\kappa$ to the eikonal equation derived from singular perturbation theory in the low-diffusion limit~\cite{tyson-keener-1988,lewis-kareiva-1993,volpert-petrovskii-2009} is in fact an exact, nonperturbative contribution originating from diffusion occurring transversally, i.e., along the wave front. For fronts moving inward into an empty domain, we identify three phases of the solution: an establishment phase, strongly influenced by conditions on the domain boundary, a travelling-wave-like phase, well described by the eikonal equation $v=c-D\kappa$ predicted by singular perturbation theory in the low-diffusion limit, and a dip-filling phase, in which the eikonal equation possesses further curvature dependences arising from the collision and interaction of inward-moving fronts. At large curvature, the normal velocity of colliding circular fronts is proportional to curvature, with an additional contribution $-D\kappa$ originating from diffusion occurring normally to wave fronts, and a time-dependent contribution originating from nonconservative changes to the solution described by the reaction term. The latter contribution becomes time-independent if $D=0$ for any choice of the reaction term $F(u)$. These behaviours are confirmed by the explicit long-time solutions of Skellam's model ($F(u)=\lambda u$), the saturated growth model ($F(u)=\lambda (1-u)$), as well as the strong Allee source term ($F(u)=\lambda u(u-a)(1-u)$), which is a common model for bistable populations~\cite{lewis-kareiva-1993,Taylor2005,Li2022}.

Our numerical simulations suggest that for a broad range of domains and reaction terms $F(u)$, inward-moving fronts that meet and interact evolve into colliding circular fronts. This behaviour is expected to be due initially to the exact linear dependence of $v$ upon $\kappa$ that arises from transverse diffusion, and the fact that mean curvature flows evolve interfaces into shrinking circles~\cite{brakke-1978,grayson-1987,chou-zhu-1999,liu-2013}. However, this behaviour likely depends on the strength of this curvature-dependent term relative to other contributions to the normal velocity. There are situations in reaction--diffusion systems with degenerate nonlinear diffusion where inward-moving fronts do not round off as circles before disappearing~\cite{mccue-etal-2019}, in which case other expressions for the eikonal equation may hold.  Excitable systems comprising several reaction--diffusion equations can present more complex wave patterns, such as spiral waves and wave trains~\cite{murray-2}. \cbeg Calcium waves and other reaction--diffusion waves may also evolve on the surface of curved domains~\cite{cheer-etal-1987,maselko-showalter-1989}. In these cases, the curvature of the surface also influences travelling wave speeds~\cite{grindrod-lewis-murray-1991,bialecki-etal-2020}. It is unclear how our results apply to these situations. Generalising the methods presented in this paper to more general diffusion operators including the Laplace--Beltrami operator could be the subject of future investigations. \cend Another situation of interest is the investigation of curvature-dependent front speeds in metastable systems with wave-pinning, such as the Allen--Cahn equation with $F(u)=u(1-u^2)$~\cite{Westdickenberg2021}, where the dip-filling phase is absent. Finally, generalisations of our analyses to heterogeneous excitable media~\cite{berestycki-hamel}, and to Fisher--Stefan models representing reaction--diffusion systems coupled with moving boundary conditions~\cite{ElHachem2019,mccue-elhachem-simpson-2021} are also of interest, both in situations supporting travelling-wave solutions as well as in dip-filling situations.
% Open questions: Unclear if there are situations (sources F(u)) for which fronts don't become circular in the long-time limit: see McCue rectangular aspect ratio. Fisher-KPP ~ Skellam not trivial, suggests scaling properties, RG

\subsection*{Acknowledgments}
This research was supported by the Australian Research Council (DP180101797, DP190102545).

\begin{appendices}
\section{Normal and transverse components of the Laplacian}\label{appx:laplacian}
In this appendix, we decompose the Laplacian into a component normal to level sets of $u$ and transverse components tangent to level sets of $u$, and show that the transverse components are proportional to the mean curvature $\kappa = \div\b n$. The mean curvature here is defined as the sum of principal curvatures, rather than their average~\cite{sethian-1999,giga-2006}, and $\b n = -\grad u/|\grad u|$ is the unit normal vector to level sets of $u$ pointing toward decreasing values of $u$. We start by expanding the divergence of $\b n$ using standard rules of differentiation:
\begin{align}\label{kappa-appx}
    \kappa = \div\b n = -\div\left(\frac{\grad u}{|\grad u|}\right) = -\frac{\laplacian u}{|\grad u|} -\grad u\vdot \grad\left(\frac{1}{|\grad u|}\right) = -\frac{\laplacian u}{|\grad u|} + \grad u \vdot \frac{\grad{\left(|\grad u|\right)}}{|\grad u|^2}.
\end{align}        
Furthermore,
\begin{align}\label{grad-hess}
    \grad\left(|\grad u|\right) = \grad\sqrt{\grad u \vdot \grad u} = \frac{(\grad{\grad u})^T\grad u + \grad u^T\vdot \grad{\grad u}}{2|\grad u|} = \frac{H(u)\grad u}{|\grad u|},
\end{align}
where
\begin{align*}
  H(u)=\grad{\grad u}=\left\{\frac{\p^2u}{\p x_i\p x_j}\right\}_{i,j=1}^m
\end{align*}
is the Hessian matrix collecting all second partials of $u$ in $m$ dimensions of space ($m=2,3$). We thus have, from Eqs~\eqref{kappa-appx} and~\eqref{grad-hess}:
\begin{align}\label{kappa-normal-transverse}
    \kappa = -\frac{\laplacian u}{|\grad u|} + \frac{1}{|\grad u|^3}\grad u\vdot H(u)\grad u = -\frac{1}{|\grad u|}\left(\laplacian u - \b n^\text{T} H(u) \b n\right).
\end{align}
By defining the derivative in the normal direction $\p/\p n$ as $\p/\p n=\b n\vdot \grad$, we also see that
\begin{align*}
    u_{nn} = \frac{\p^2u}{\p n^2} = \b n\vdot \grad{\left(\b n\vdot \grad u\right)} = - \b n\vdot \grad\left(|\grad u|\right) = \b n^\text{T} H(u)\b n,
\end{align*}
where we used $\b n\vdot\grad u = -|\grad u|$ and Eq.~\eqref{grad-hess}. Reorganising Eq.~\eqref{kappa-normal-transverse}, the Laplacian can therefore be decomposed into a normal component $u_{nn}=\b n^\text{T}H(u)\b n$, and a transverse component proportional to the mean curvature:
\begin{align}\label{laplacian-normal-transverse}
    \laplacian u = u_{nn} - \kappa |\grad u|.  
\end{align}

\section{Computer code}
Key algorithms used to generate results are freely available on Github at \url{https://github.com/prbuen/Buenzli2022_CurvatureReacDiff}
\end{appendices}

\end{document}